%% file: main.tex
\documentclass[lettersize,journal]{IEEEtran}
\usepackage{amsmath,amsfonts}
\usepackage{algorithmic}
\usepackage{algorithm}
\usepackage{array}
\usepackage{textcomp}
\usepackage{stfloats}
\usepackage{url}
\usepackage{verbatim}
\usepackage{graphicx}
\usepackage{subcaption}
\usepackage{cite}
\usepackage{xcolor}
\usepackage{xspace}

\newcommand{\sysname}{$\textit{CoUAV-Pro}$\xspace}

\newcommand*{\shen}[1]{\textcolor{black}{#1}}
\begin{document}

\title{Collaborative UAVs Multi-task Video Processing Optimization Based on Enhanced Distributed Actor-Critic Networks }


\author{\IEEEauthorblockN{Ziqi Rong\textbf{\textsuperscript{*}},
Qiushi Zheng\textbf{\textsuperscript{*}},~\IEEEmembership{Member,~IEEE}\thanks{\textbf{*}Both authors contributed equally to this work. (Corresponding author: Zhishu Shen and Jiong Jin)},
Zhishu Shen,~\IEEEmembership{Member,~IEEE,} 
Xiaolong Li,
Tiehua Zhang,~\IEEEmembership{Member,~IEEE,}
Zheng Lei,
and Jiong~Jin~\IEEEmembership{Member,~IEEE,}}

\thanks{Ziqi Rong, Zhishu Shen and Xiaolong Li are with the School of Computer Science and Artificial Intelligence, Wuhan University of Technology, Wuhan, China (e-mail: rongziqi@whut.edu.cn, z\_shen@ieee.org, m2002lixiaolong@whut.edu.cn).}
\thanks{Qiushi Zheng and Jiong Jin are with the School of Science, Computing and Engineering Technologies, Swinburne University of Technology, Melbourne, Australia (e-mail: \{qiushizheng, jiongjin\}@swin.edu.au).}
\thanks{Zheng Lei is with the School of Engineering, Swinburne University of Technology, Melbourne, Australia (e-mail: zlei@swin.edu.au)}
\thanks{Tiehua Zhang is with the School of Computer Science and Technology, Tongji University, Shanghai, China (e-mail:  tiehuaz@tongji.edu.cn).}

}

\markboth{Journal of \LaTeX\ Class Files,~Vol.~14, No.~8, August~2021}%
{Shell \MakeLowercase{\textit{et al.}}: A Sample Article Using IEEEtran.cls for IEEE Journals}


\maketitle

\begin{abstract} 
With the rapid advancement of the Internet of Things (IoT) and Artificial Intelligence (AI), intelligent information services are being increasingly integrated across various sectors, including healthcare, industry, and transportation. Traditional solutions rely on centralized cloud processing, which encounters considerable challenges in fulfilling the Quality of Service (QoS) requirements of Computer Vision (CV) tasks generated in the resource-constrained infrastructure-less environments. In this paper, we introduce a distributed framework called \sysname for multi-task video processing powered by Unmanned Aerial Vehicles (UAVs). This framework empowers multiple UAVs to meet the service demands of various computer vision (CV) tasks in infrastructure-less environments, thereby eliminating the need for centralized processing. Specifically, we develop a novel task allocation algorithm that leverages enhanced distributed actor-critic networks within \sysname, aiming to optimize task processing efficiency while contending with constraints associated with UAV's energy, computational, and communication resources. Comprehensive experiments demonstrate that our proposed solution achieves satisfactory performance levels against those of centralized methods across key metrics including task acquisition rates, task latency, and energy consumption. 
\end{abstract}


\input{1_introduction}
\input{2_relatedwork}
\input{3_model}
\input{4_formulation}

\input{5_algorithm}
\input{6_experiment}
\input{7_conclusion}




%

\bibliographystyle{ieeetr}
\bibliography{ref.bib}

\vfill

\end{document}

%% file: 1_introduction.tex
\section{Introduction}\label{sec:introduction}

\IEEEPARstart{T}{he} development of the Internet of Things (IoT) and 
Artificial Intelligence (AI) technologies facilitate the deployment of diverse intelligent information services in diverse areas. These intelligent information services are essential not only in densely populated smart cities but also for a range of applications in infrastructure-less environments where no cellular network is the norm such as rural/remote areas~\cite{ShenCSUR23}. However, geographic and economic constraints limit the coverage of terrestrial networks in these vast rural/remote areas, making it challenging to guarantee satisfactory Quality of Service (QoS) for various intelligent information services~\cite{WangJSAC21,LiuNW22,ZhengTPDS24}. The non-terrestrial networks utilizing aerial components, such as Unmanned Aerial Vehicle (UAV) and Low-Earth Orbit (LEO) satellites, are regarded as a promising solution to extend the connectivity beyond current communication networks~\cite{XiaoJSAC24,MahboobCST24}. 

Recently, extensive research has explored the benefits of integrating computing capabilities into Unmanned Aerial Vehicles (UAVs) as a means to provide intelligent information services in infrastructure-less environments. For instance, a UAV equipped with onboard sensors and cameras, combined with AI inference models, can support a variety of intelligent services, including smart transportation and environmental monitoring~\cite{McEnroeIoT22,DaiTMC24}. Meanwhile, many IoT devices deployed in infrastructure-less environments are energy-constrained and lack the capacity for long-distance communication or computation-intensive intelligence analysis. In this context, UAVs can not only serve as relay nodes to transmit data to distant ground Base Stations (BSs) but also perform various data processing tasks, thereby reducing both data processing and communication latency~\cite{NingJSAC21,QuJSAC21,HaoTMC24}. 

The integration of UAVs with real-time Computer Vision (CV) capabilities further enhances their potential to enable autonomous decision-making and task completion in rural settings. By leveraging onboard cameras and sensors, UAVs can perform CV tasks such as object detection, scene recognition, and tracking, which are crucial for a variety of applications like agricultural monitoring, wildlife tracking, and disaster response~\cite{ChenDrones23}. These capabilities allow UAVs to analyze visual data locally, reducing the reliance on distant cloud computing resources and minimizing communication delays that could otherwise hinder real-time decision-making. Moreover, the incorporation of AI-based CV algorithms enables UAVs to adapt to dynamic environments and changing conditions in real-time~\cite{WuGRSM22}. For instance, in agricultural applications, UAVs can autonomously detect crop diseases or pests from aerial images~\cite{SangaiahTNSE24}, make decisions about targeted pesticide spraying, and adjust their flight paths accordingly. Similarly, in environmental monitoring, UAVs can identify irregularities such as forest fires, ensuring timely interventions~\cite{LiChangIOTJ24}.

Despite these promising achievements, several challenges remain in integrating UAVs into IoT-based intelligent CV services, especially in infrastructure-less environments. UAVs are constrained by limited battery capacity, which affects both their flight endurance and computational power. Efficiently offloading computation-intensive tasks to ground BSs or other UAVs, while minimizing energy consumption and maintaining low processing latency, is a key research problem~\cite{ZhaoTWC22}. Task allocation strategies must adapt to the dynamic nature of infrastructure-less environments and the varying complexity of tasks~\cite{ChenTVT24}. Designing intelligent and adaptable task allocation mechanisms is crucial for maximizing the use of computational resources, improving overall system performance, and fulfilling quality of service (QoS) requirements. \shen{The multi-task processing in infrastructure-less environments also face information synchronization issues arise from  the high-latency and costly satellite communication. This not only hinders real-time coordination between UAVs and ground stations, but also results in substantial communication costs to maintain synchronization across distributed systems,  ultimately diminishing system efficiency.}  Additionally, UAV trajectory planning becomes more complex in infrastructure-less environments due to varying terrain and sparse communication infrastructure~\cite{ShuklaCST24}. UAVs must ensure coverage of target areas, optimize their flight paths to conserve energy, and avoid obstacles, all while staying within their operational limits.


To address the aforementioned challenges, we propose a collaborative UAVs multi-task video processing \sysname that maximizes the task collection rate and the task processing rate energy-efficiently. This framework is designed to facilitate multitask video collection and processing across various multiple UAVs. Specifically, 
\shen{it leverages a distributed algorithm to overcome the information synchronization issues inherent in infrastructure-less environments, enabling individual ground stations to assign multi-tasks independently and facilitating periodic model synchronization and communication among UAVs.} By implementing periodic model synchronization among agents and enabling communication between colliding UAVs, \sysname can minimize the assignment of duplicate tasks to UAVs in subsequent rounds. This collaborative strategy can approach the performance metrics of centralized systems while effectively managing the constraints inherent in infrastructure-less environments. The main contributions of this paper are summarized below:\begin{itemize}
    \item  We design a distributed multi-task video processing framework \sysname for infrastructure-less environments. It enables efficient task allocation among multiple UAVs and facilitates information sharing during the data collection process, thereby enhancing service reliability and performance without the need for continuous reliance on a central server.
    \item We formulate the optimization problem as a Selective Multiple Depot Multiple Traveling Salesman Problem (SMD-MTSP). To reduce computational complexity, we propose an efficient task allocation algorithm based on enhanced distributed actor-critic networks. This algorithm leverages collaborative UAVs to facilitate task allocation in resource-constrained environments while achieving satisfactory data processing performance.
    \item We conduct extensive experiments to evaluate the performance of \sysname in three types of video processing and computational tasks. The comprehensive experimental results demonstrate that our distributed method can attain comparable performance as a centralized method in terms of various criteria including task collection rate, task completion rate, task processing time, UAV utilization rate, and energy consumption.

\end{itemize}

The remainder of this paper is organized as follows: 
Section~\ref{sec:relatedwork} summarizes the related work.  Section~\ref{sec:model} introduces the system model, followed by the problem formulation in Section~\ref{sec:formulation}. Our proposed \sysname is detailed in Section~\ref{sec:algorithm}. Section~\ref{sec:evaluation} validates the effectiveness of  \sysname against comparative methods. Section~\ref{sec:conclusion} summarizes this work.

%% file: 2_relatedwork.tex
\section{Related Work}\label{sec:relatedwork}

\subsection{UAV-based MEC Networks}
The integration of Multi-access Edge Computing (MEC) into UAV systems has garnered significant research interest due to its potential to enhance computational efficiency and minimize latency~\cite{TaoJSAC24}. A critical focus area is task offloading, which involves distributing computational workloads from UAVs to edge servers. Tian \textit{et al.} proposed a novel user satisfaction model that incorporates both task delay and energy savings, introducing a genetic algorithm-based solution to optimize task offloading and UAV scheduling~\cite{TianTWC23}. Similarly, Kang \textit{et al.} explored a joint optimization framework employing a Multi-Agent Proximal Policy Optimization (MAPPO)-based algorithm to enhance task offloading and resource management~\cite{KangIOTJ23}. Lin \textit{et al.} developed an algorithm integrating Deep Reinforcement Learning (DRL) with linear programming to optimize UAV flight time, three-dimensional trajectories, and binary offloading decisions~\cite{LinTWC23}. Additionally, Nguyen \textit{et al.} presented a collaborative approach for UAV-assisted MEC networks, addressing the challenges of limited computational capacity and battery life of individual UAVs serving multiple users~\cite{NguyenIOTJ23}.

Furthermore, trajectory planning is a crucial component of UAV-assisted MEC systems. Current research in this area focuses on designing optimal flight trajectories that ensure comprehensive area coverage while optimizing energy efficiency, reducing latency, and maintaining reliable communication links between UAVs and ground stations. Ejaz \textit{et al.} introduced a deep Q-learning-based framework for multi-UAV systems aimed at QoS and route planning\cite{EjazTNSM24}. Liu \textit{et al.} concentrated on UAV trajectory design and resource allocation within a Heterogeneous Mobile-Edge Computing (HMEC) system, where UAVs provide support to energy-limited IoT devices via wireless power transfer~\cite{LiuIOTJ24}. Li \textit{et al.} investigated UAV trajectory design in multi-UAV-assisted MEC networks, developing a robust optimization framework that minimizes energy consumption by jointly optimizing UAV trajectories, task partitioning, and resource allocation~\cite{LiIOTJ24}. However, most of the proposed methods overlook the dynamic interactions between UAVs, leading to inefficiencies in task scheduling and resource management.

\subsection{UAV-based Video Processing Methods}
With the advancement of deep learning technologies, UAVs equipped with real-time object detection algorithms, such as You Only Look Once (YOLO) and OpenCV, have demonstrated significant potential in various applications, including surveillance~\cite{ZhengIoT24}, disaster response~\cite{LiIoT24}, and precision agriculture~\cite{SangaiahTNSE24}. Current research emphasizes adapting these algorithms to the limited computational capacity of UAVs while ensuring high detection accuracy and real-time performance~\cite{ChenDrones23}. For instance, Chen \textit{et al.} proposed an adaptive bitrate video delivering strategy in UAV-assisted MEC networks, formulating the problem of joint cache placement and video delivery scheduling to minimize the total expected system latency with energy consumption constrained~\cite{ChenTMC24}.

Given the limitations of traditional terrestrial communication networks in supporting IoT devices deployed in isolated areas, innovative strategies are essential to improve connectivity and resource availability. In this regard, Qin \textit{et al.} proposed a multiple UAV-assisted air-ground collaborative edge computing network model designed to minimize energy consumption for IoT devices located in remote areas, utilizing deep reinforcement learning for task offloading and trajectory optimization~\cite{QinTNSE24}. Furthermore, Space-Air-Ground Integrated Networks (SAGINs) have emerged as promising architecture for providing continuous, large-scale surveillance and real-time data processing~\cite{ShenCSUR23}. These networks integrate satellite, UAV, and terrestrial infrastructures, extending the coverage of the existing terrestrial networks. Yang \textit{et al.} proposed a Quality of Experience (QoE)-oriented transmission scheduling policy for UAV-assisted video streaming in satellite-terrestrial networks, employing a cache-constrained Markov decision process in conjunction with a hybrid reinforcement learning algorithm~\cite{YangICC24}. By implementing efficient information transfer mechanisms, UAVs can exchange status, task, and environmental data in real-time, allowing for improved adaptation to complex and dynamic scenarios.

%% file: 3_model.tex
\section{System Model}\label{sec:model}


\begin{figure}[tb!]
    \centering
    \includegraphics[width=\columnwidth,trim=0 100 0 100,clip]{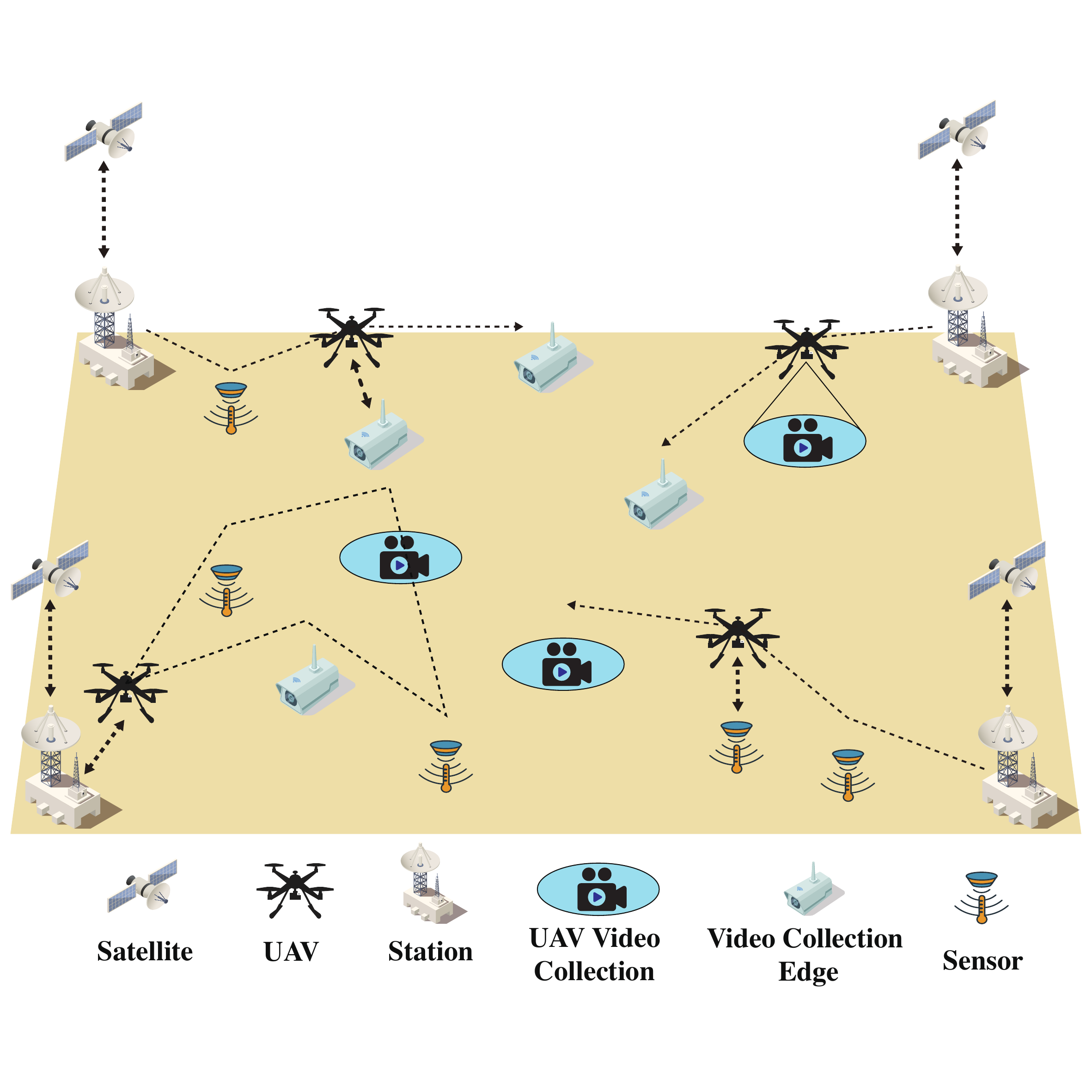} 
    \caption{Overview of the network model. }
    \label{fig:overview}
\end{figure}

\subsection{Network Model}~\label{subsec:nwmodel}
As shown in \figurename~\ref{fig:overview}, we assume an aerial computing network model that mainly incorporates multiple \textbf{UAVs} and \textbf{ground nodes} to perform comprehensive data processing from information collection to decision-making in the infrastructure-less environments. The ground node $n \in \mathcal N$ includes \textbf{ground edge} and \textbf{ground IoT sensors}. Each UAV $ u \in \mathcal U$, 
serves as a mobile data processor and aggregator for the given tasks. Each UAV $u$ is assigned to a \textbf{ground station} $g \in \mathcal G$. 
The ground station $s$ is located outside the mission area and serves as a launch and return point for a specific subset of UAVs. Each $s$ supports UAVs by providing resources such as battery recharging, data storage, and task deployment prior to each mission. Upon mission completion, UAVs return to their assigned station for data offloading and necessary maintenance.

Each UAV $u$ departs from its original station, navigates to a designated grid region within an area of \( A \times A \), collects and processes tasks, and then returns to its station. It is worth noting that the collaboration between UAVs is essential for efficient mission planning, which enables multiple UAVs to intelligently coordinate flight paths and maximize mission completion. Upon returning to a ground station, UAV $u$ unloads the collected tasks for preprocessing, and satellite communication is used to periodically upload the processed data.

The \textbf{tasks} \( {Tk} = \{Tk_1, Tk_2, \dots, Tk_m\} \) are divided into numerous categories, where the task belongs to the category \( CT \) denoted as \( Tk_m^{CT} \). In this paper, \( Tk \) include three primary types: \textbf{UAV video collection} \( Tk_m^{1} \), \textbf{edge video collection} \( Tk_m^{2} \), and \textbf{sensor environmental data collection} \( Tk_m^{3} \) as illustrated in \figurename~\ref{fig:overview}. UAVs equipped with onboard cameras capture real-time video footage from optimal altitudes, providing valuable situational data. Ground edge nodes, which also capture video, offer continuous coverage of specific locations, enabling fixed-point monitoring in critical areas. Additionally, environmental sensors scattered across the region collect vital data on atmospheric conditions, such as temperature, humidity, and air quality. These tasks are essential in infrastructure-less environments, where continuous monitoring and real-time data aggregation provide the foundation for mission success. Let the priority of each task type be \( Pr_m^{CT} \), where \( {Pr}_m^{1} > {Pr}_m^{2} > {Pr}_m^{3} \). UAVs serve as both data collectors and processors, effectively reducing communication latency and facilitating multi-task video processing in various scenarios like disaster response and environmental monitoring, where timely and adaptive decision-making is essential.

\subsection{Communication Model}
\subsubsection{UAV to ground node} 
The communication between UAV $i$ and ground node $n$ is modeled by considering the data transmission rate, free-space path loss, probability of Line-of-Sight (LoS), channel gain, and Signal-to-Noise Ratio (SNR). The UAVs communicate with ground edge nodes distributed randomly across an \( A \times A \) region. The data transmission rate \( R_{i,n}^{U2G} \) between UAV \( i\) and ground node \( n \) is given by the Shannon capacity formula:
\begin{equation}
R_{i,n}^{U2G}= B^G \log_2 \left( 1 + \frac{P_i h_{i,n}}{N^{U2G}_0 B^G} \right)
\end{equation}
where \( B^G \) is the channel bandwidth, \( P_i \) is the transmit power of UAV \( i\), \( h_{i,n} \) is the channel gain between UAV \( i \) and ground node \( n \), and \( N_0 \) is the noise power spectral density. The channel gain \( h_{i,n} \) is a function of the free-space path loss and the probability of LoS between UAV $i$ and the ground node $n$.

The free-space path loss \( L^{U2G}_{i,n} \) is expressed as:
\begin{equation}
L^{U2G}_{i,n} = \left( \frac{4 \pi f_c d_{i,n}}{c} \right)^2
\end{equation}
where \( f_c \) is the carrier frequency, \( d_{i,n} \) is the Euclidean distance between the UAV and the ground node, and \( c \) is the speed of light. The probability of LoS \( P_{\text{LoS}}(h_{i,n}) \) is modeled based on the elevation angle \( \theta_{in} \) between the UAV and the ground node as follows: 
\begin{equation}
P^{U2G}_{\text{LoS}}(h_{i,n}) = \frac{1}{1 + a \exp \left( -b \left( \theta_{in} - a \right) \right)}
\end{equation}
where \( a \) and \( b \) are constants depending on the environment. 
The channel gain \( h_{i,n} \) is then modeled as a combination of LoS and Non-Line-of-Sight (NLoS) conditions:
\begin{equation}
h_{i,n} = \frac{P_{\text{LoS}}(h_{i,n}) h_{\text{LoS}} + (1 - P_{\text{LoS}}(h_{i,n})) h_{\text{NLoS}}}{L^{U2G}_{i,n}}
\end{equation}
where \( h_{\text{LoS}} \) and \( h_{\text{NLoS}} \) are the channel gains under LoS and NLoS conditions, respectively.



\subsubsection{UAV to UAV} 
The communication between UAVs operates under free-space path loss conditions, and the data rate \( R \) between UAV $i$ and $j$ can be derived as:
\begin{equation}
R_{i,j}^{U2U} = B^U \cdot \log_2 \left(1 + \frac{P_i h_{i,j}}{N^{U2U}_0 B^U}\right)
\end{equation}
where \( B^U \) denotes the channel bandwidth. The transmission power \( P_i \) of the sending UAV determines the energy level of the signal sent to the receiving UAV. Antenna gain values \( G_i \) and \( G_j \) refer to the efficiencies of the transmitting and receiving UAV antennas, respectively, with each value being a function of the antennas’ orientations and designs. The wavelength \( \lambda \) is associated with the communication signal’s frequency \( f \) and can be calculated as \( \lambda = \frac{c}{f} \), where \( c \) is the speed of light. The distance \( d \) between the two UAVs affects the signal strength as it propagates, with greater distances generally reducing signal strength under free-space conditions. \( N_0 \) represents the noise power spectral density, which captures the environmental noise level that can interfere with the received signal.



\subsection{Computation Model}

The total computational load for task $m$, given that only a fraction \( \beta_m \) of the task is processed onboard, is described in terms of the required number of floating-point operations (FLOPs) as follows:
\begin{equation}
C_m^{\text{FLOPS}} = \beta_m \sum_{k=1}^{K} x_{k,m} \cdot \gamma_{k} S_m
\end{equation}
where \( \beta_m \) represents the fraction of task $m$ that is processed onboard, constrained within the range \( 0 \leq \beta_m \leq 1 \). This parameter allows for adaptive allocation of computational resources, enabling flexibility in determining the proportion of the task handled directly by the UAV. The binary variable \( x_{k,m} \) serves as an indicator for the task category $Tk_m^{CT}$, where \( x_{k,m} = 1 \) if task $m$ belongs to category \( k \), and \( x_{k,m} = 0 \) otherwise. The category-specific parameter \( \gamma_m \) acts as a conversion factor, translating the data size of task $m$, specific to category \( k \), into the corresponding number of FLOPs required for processing. \( S_m \) denotes the size of task $m$, capturing the total data volume associated with the onboard processing workload.

The computation capacity of a multi-core system, such as a Raspberry Pi 4 used in a UAV, is determined by the processing power across its active cores. The total processing capacity denoted by \( f_{\text{total}} \) can be expressed as:
\begin{equation}
f_{\text{total}} = N_{\text{cores}} \cdot f_{\text{CPU}} \cdot \eta_f
\end{equation}
where \( N_{\text{cores}} \) is the number of active CPU cores, \( f_{\text{CPU}} \) is the processor's operating frequency in cycles per second, and \(\eta_f\) represents the number of FLOPs each core can perform per cycle. 

The computation delay \( D_m \) for task $m$ is defined as the time needed to complete its required operations, adjusted for system overhead. It can be expressed as:
\begin{equation}
D_m = \frac{C_m^{\text{FLOPS}}}{f_{\text{total}}} + \tau_m 
\end{equation}
where the term \( \tau_m \) captures system overhead, accounting for delays due to scheduling, memory access, and I/O operations.

\subsection{Mobility of UAV}
The distance between the UAV's current position, represented by $(x(t), y(t), z(t))$, and the location of a specific task $i$, represented by $(x_i, y_i, z_i)$, can be divided into two components: the horizontal distance $d_{\text{horizontal}}$ and the vertical distance $d_{\text{vertical}}$. The total Euclidean distance $d_{\text{U2t}}$ between the UAV and the task location is represented by a combined expression that separately accounts for the horizontal and vertical distance components:
\begin{equation}
\begin{split}
d_{\text{U2t}}=\sqrt{(x_i - x(t))^2 + (y_i - y(t))^2} + \sqrt{(z_i - z(t))^2}
\end{split}
\end{equation}

In this paper, we ignore the additional or reduced energy consumption resulting from UAV acceleration and deceleration, which is reasonable in scenarios when the duration of UAV maneuvering constitutes only a minor  portion of the overall operation time~\cite{zeng2019energy}. The closed-form power consumption models for a multi-rotor UAV in horizontal flight and vertical flight have been created, including hovering, cruising, ascent and descent~\cite{gong2023modelling}. When a UAV hovers in place, such as during video data collection, it requires a power \( P_{\text{hover}} \) to maintain its position. This hovering power, \( P_{\text{hover}} \), consists of two components: the blade profile power \( P_{bl} \) and the induced power \( P_{in} \) needed to sustain the UAV in a stationary position:
\begin{equation}
\begin{split}
P_{\text{hover}} &= P_{bl} + P_{in} \\& =\frac{W^{3/2}}{\sqrt{n_{r} \rho  A_{\text{rotor}} }} C_T^{-3/2} \frac{\delta}{8} s + \frac{W^{3/2}}{\sqrt{2 n_{r} \rho  A_{\text{rotor}} }} (1 + k)
\end{split}
\end{equation}
where \( W \) denotes the weight of the UAV with \( n_{r} \) rotors, \( \eta \) represents the efficiency factor, \( \rho \) is the air density, \(s \) is rotor solidity, and \( A_{\text{rotor}} \) is the rotor disk area in square meters. The parameter \( C_T \) is the thrust coefficient, \( \delta \) is the profile drag coefficient, and \( k \) is an incremental correction factor for induced power.

For extended movement across a target area, cruising power \( P_{\text{cruise}} \) is required:
\begin{equation}
\begin{split}
P_{\text{cruise}} & = P_{bl} + \frac{3}{8} \delta \sqrt{\frac{W n_{r} \rho A}{C_T}} s V^2 \\&+ P_{in} \left( \sqrt{1 + \frac{V^4}{4 v_0^4}}- \frac{V^2}{2 v_0^2} \right)^{1/2} + \frac{n_{r}}{2} S_{FP||} \rho V^3
\end{split}
\end{equation}
where \( v_0 \) is the mean rotor induced velocity in hover, \( S_{FP||} \)is the fuselage equivalent flat plate area in the horizontal orientation. Additionally, \( V \) represents a constant forward speed of the UAV when it is flying.


For tasks involving altitude adjustments, such as ascending or descending to collect data at specific heights, the additional vertical power needed for the multi-rotor UAV can be calculated as follows:
\begin{equation}
\begin{split}
P_{\text{ascent}} & = \frac{1}{2} W V + \frac{n_{r}}{4} S_{FP \perp} \rho V^3 
\\& + \left( \frac{W}{2} + \frac{n_{r}}{4} S_{FP \perp} \rho V^2 \right) \sqrt{\left(1 + \frac{S_{FP \perp}}{A}\right) V^2 + \frac{2W}{n_{r} \rho A}}
\end{split}
\end{equation}

\begin{equation}
\begin{split}
P_{\text{descent}} & = \frac{1}{2} W V - \frac{n_{r}}{4} S_{F_{P_\perp}} \rho V^3 \\&
+ \left( \frac{W}{2} - \frac{n_{r}}{4} S_{F_{P_\perp}} \rho V^2 \right) \sqrt{\left( 1 - \frac{S_{F_{P_\perp}}}{A} \right) V^2 + \frac{2 W}{n_{r} \rho A}}
\end{split}
\end{equation}
where \(S_{FP \perp}\) represents the fuselage equivalent flat plate area in the vertical status.



\subsection{Task Processing Model}~\label{sec:tpm}


The task processing model addresses the energy consumption and power requirements for three distinct operational scenarios of a UAV, each involving different task collection maneuvers. The power calculations incorporate adjustments in speed and altitude that directly impact the energy demands for hovering, cruising, acceleration, deceleration, and altitude changes. The descent and ascent altitudes, \( z_{\text{descend}} \) and \( z_{\text{ascend}} \), are defined using the maximum and minimum altitudes \( z_{\text{max}} \) and \( z_{\text{min}} \), \( z_{\text{descend}} = |z_{\text{max}} - z_{\text{min}}| \) and \( z_{\text{ascend}} = |z_{\text{min}} - z_{\text{max}}| \) separately. Therefore, the \( t_{\text{descend}} \) and \( t_{\text{ascend}} \) can be calculated by \(\frac{z_{\text{descend}} }{v_{\text{descend}}}\) and \(\frac{z_{\text{ascend}} }{v_{\text{ascend}}}\). The following models quantify the power consumption for each task scenario, considering the UAV's flight profile and optimizing energy efficiency for task collection.

\textbf{UAV Video Collection}: In this model, the UAV reduces its speed to a lower cruise speed, \( v_{\text{cruise, low}} \), and descends to a designated altitude for task collection. During the descent and the subsequent ascent back to its original altitude after completing the task, the altitude-related energy consumption is represented by \( P_{\text{descent}} \) and \( P_{\text{ascent}} \), respectively. The time required for video collection is denoted as \( t_{Tk_m^{1}} \). The total energy requirement for UAV video collection denoted as \( E_{{Tk_m^{1}}} \), is therefore given by:
\begin{equation}
E_{{Tk_m^{1}}} = P_{\text{descent}} \cdot t_{\text{descend}}+ P_{\text{cruise, low}} \cdot t_{Tk_m^{1}} + P_{\text{ascent}} \cdot t_{\text{ascend}}
\end{equation}

\textbf{Edge Video Collection}: In this scenario, the UAV decelerates from its cruise speed to a complete stop, hovering above the task location to begin collection. It then descends to the ground to collect task \( m \). Once the task is collected, with its initial size noted as \( s_m \), the UAV ascends to its previous altitude, before accelerating to resume its original cruise speed. The total energy required for edge video collection denoted as \( E_{{Tk_m^{2}}} \), is therefore calculated as follows:
\begin{equation}
E_{{Tk_m^{2}}} = P_{\text{descent}} \cdot t_{\text{descend}} + P_{\text{hover}} \cdot \frac{s_m}{R_{i,m}^{U2G}} + P_{\text{ascent}} \cdot t_{\text{ascent}}
\end{equation}

\textbf{Sensor Environmental Data Collection}: This task model involves reducing the UAV’s altitude for task collection without adjusting its cruise speed. The power consumption for this model includes only the altitude change and cruise power during task collection \( t_{Tk_m^{3}} \). After the task is collected, the UAV ascends back to the original altitude. The total power required for this task model, \( P_{\text{model 3}} \), is represented as:
\begin{equation}
E_{{Tk_m^{3}}} = P_{\text{descent}} \cdot t_{\text{descend}}+ P_{\text{cruise}} \cdot t_{Tk_m^{3}} + P_{\text{ascent}} \cdot t_{\text{ascend}}
\end{equation}


%% file: 4_formulation.tex
\section{Problem Formulation}~\label{sec:formulation}

\subsection{Flight Energy Constraint}

To ensure reliable task completion and safe return, the UAV's movement is governed by an energy constraint linked to its battery capacity. The UAV's total travel distance consists of three distinct segments: (a) the initial distance from the ground station to the first task location; (b) the cumulative distance between consecutive tasks; and (c) the distance from the final task location back to the ground station. Thus, the total movement distance \( \textit{d}_i \) is defined as:
\begin{equation}
\textit{d}_i = \sum_{\forall m} x_{i,m} \cdot d_{i,m} + \sum_{\forall m} \sum_{\forall n} x_{i,m} \cdot x_{i,n} \cdot d_{m,n} + \sum_{\forall n} x_{i,n} \cdot d_{i,n}
\end{equation}
where \( d_{i,j} \) represent the distance between station of UAV \( i \) and task \( m \), and \( d_{m,n} \) the distance between tasks \( m \) and \( n \) ($m \neq n$). \( x_{i,m} \) is a binary decision variable, where \( x_{i,m} = 1 \) if UAV \( i \) is assigned to visit task \( m \), and \( x_{i,m} = 0 \) otherwise.

The total energy consumption must not exceed the UAV \(i\) battery capacity \( B_{\text{flight},i} \) to ensure mission completion and return capability. The flight energy constraint is therefore expressed as:
\begin{equation} \label{eq:lim1}
\frac{\textit{d}_{i} P_{\text{cruise}}}{v_{\text{cruise}}} + \sum_{\forall i} \sum_{\forall CT} E_{{\textit{Tk}_m^{CT}}} \cdot x_{m,\textit{TK}} \cdot x_{i,m}\leq B_{\text{flight},i}
\end{equation}

\subsection{Processing Energy Constraint}
Each UAV’s processing capability is constrained by its limited battery capacity designated for task processing. The total energy consumed to process tasks must not exceed this limit, as expressed by:
\begin{equation}\label{eq:lim2}
\sum_{\forall m} x_{i,m} \cdot s_m \cdot p_{i,m} \cdot e_{\text{process}} \leq B_{\text{process},i}, \quad \forall i 
\end{equation}
where $p_{i,m}$ is the proportion of task $m$ processed by UAV $i$, $e_{\text{process}}$ represents the energy consumption per unit of task size, and $B_{\text{process},i}$ is the processing energy capacity of UAV $i$. The storage constraint ensures that each UAV maintains within its storage capacity both the initial size of collected tasks and the reduction in size from any partially processed tasks, which is represented by:
\begin{equation} \label{eq:lim3}
\sum_{\forall m} x_{i,m} \cdot (s_m - p_{i,m} \cdot s_m) \leq S_i, \quad \forall i 
\end{equation}
where $S_i$ is the total storage capacity of UAV $i$. The following assignment constraint is applied to ensure that each task is collected by only one UAV:
\begin{equation} \label{eq:lim4}
\sum_{\forall i}x_{i,m} \leq 1, \quad \forall m 
\end{equation}

\subsection{Objective Function}
For all UAV $i$, the objective of the optimization problem is to maximize the overall task collection and processing efficiency, which is expressed as:
\begin{equation}
\max \sum_{\forall i} \sum_{\forall m} v_m - \alpha \sum_{\forall i} E_{\text{flight},i} - \beta \sum_{\forall i} E_{\text{process},m}
\end{equation}
where 
\begin{equation}
v_m = k_m \cdot \left( 1 + r_m^\tau \right) \cdot x_{i,m}
\end{equation}

s.t.
\begin{subequations}
\begin{equation}
  \sum_{\forall i}x_{i,m} \leq 1, x_{i,m} = \{0, 1\}
\end{equation}
\begin{equation}
    p_{i,m} \leq 1
\end{equation}
\begin{equation}
   \text{Constraints}~(\ref{eq:lim1})-(\ref{eq:lim4})
\end{equation}
\end{subequations}

Here, \( v_m \) represents the value of task \( m \), \( k_m \) is a constant associated with the intrinsic value of the task, and \( r_m \) is a parameter that influences the value based on specific task characteristics. The coefficients \( \alpha \) and \( \beta \) are weighting factors that prioritize energy consumption relative to the benefits derived from task collection.

%% file: 5_algorithm.tex
\section{Multi-task Video Processing
 Using Enhanced Distributed
Actor-Critic Networks}~\label{sec:algorithm}
\subsection{Problem Optimization} 

Based on the summary in Section~\ref{sec:formulation}, this study aims to maximize the profit generated by a fleet of UAVs tasked with collecting data from various assignments, all while taking energy and storage constraints into account. To achieve this, we model the problem using a Selective Multiple
Depot Multiple Traveling Salesman Problem (SMD-MTSP) with a curved distance metric. In SMD-MTSP, given the graph $G$ = ($V$,$E$), it finds the route for the salesmen (= UAVs) starting from multiple depots (= ground stations), so that each target (= task) is visited by at most one UAV while the overall profit can be maximized. This formulation reflects the UAVs’ energy consumption in traversing curved paths between tasks $n$ and $m$, which provides a more realistic scenario for mission planning. Specifically, it can be formulated as:
\begin{equation}~\label{eq:obj}
\max \sum_{\forall i}\sum_{\forall m} (v_n \cdot x_{i,m}) - \sum_{\forall i}di
\end{equation}
s.t.
\begin{subequations}  
   \begin{equation}  \label{eqa}
       x_{i,m} \in \{0, 1\}, \quad \forall i, m 
   \end{equation}
   \begin{equation} \label{eqb}
        \sum_{\forall i} x_{i,m} \leq 1, \quad \forall m
   \end{equation}
   \begin{equation}  \label{eqc}
         \sum_{\forall m} x_{i,m} \cdot \left(E_{i,m}^{\text{move}}\right) + E_{i, \text{depot}}^{\text{move}} \leq B_{\text{flight},i}
   \end{equation}
    \begin{equation} \label{eqf}
   \sum_{\forall m} x_{i,m} \cdot \left(E_{i,m}^{\text{process}}\right)\leq B_{\text{process},i}
   \end{equation}
   \begin{equation}  \label{eqd}
         \sum_{\forall m} x_{i,m} \cdot s_m \leq S_i, \quad \forall i
   \end{equation}
   \begin{equation}  \label{eqe}
       \sum_{\forall m} x_{i,m} \cdot p_{i,m} \leq P_m, \quad \forall i
   \end{equation}
   \begin{equation}  \label{eqg}
    \sum_{\forall n} x_{mn}^i - \sum_{\forall n} x_{nm}^i = 0, \quad \forall i, m
   \end{equation}
   \begin{equation}  \label{eqh}
    \sum_{m, n \in Tk} x_{mn}^i \leq |Tk| - 1, \quad \forall Tk, i
    \end{equation}
\end{subequations}

The proposed SMD-MTSP model follows several constraints: Constraints~\ref{eqa} and \ref{eqb} indicate the assignment constraint that each task $m$ can be assigned to at most one UAV $i$. Constraint~\ref{eqb} ensures that the energy consumption \( E_i \) of each UAV \( i \) includes movement and processing energy for assigned tasks, as well as the energy required to return to the depot. Here, \( E_{i,m}^{\text{move}} = c \cdot {d}_{i,m} \) represents the energy consumed by UAV \( i \) to move to task \( m \) using the distance \( {d}_{i,m} \), where \( c \) is a scaling factor translating the distance into energy units. The term \( e_{i,m}^{\text{process}} \) denotes the energy required to process task \( m \), and \( e_{i, \text{depot}}^{\text{move}} \) is the energy needed for UAV \( i \) to return to the depot after completing its assigned tasks.

Constraint~\ref{eqd} is the storage capacity constraint, indicating that each UAV’s total assigned storage must not exceed its capacity, where \( s_m \) is the storage requirement for task \( m \), and \( S_i \) is the storage capacity of UAV \( i \). Constraint~\ref{eqe} is the processing capacity constraint that ensures that each UAV’s processing load must be within its processing capacity, i.e., each UAV can handle the processing requirements of its assigned tasks without exceeding its onboard processing capacity \( P_i \). Constraint~\ref{eqg} guarantees that each task is visited exactly once and that the route flows continuously from one location to the next, while Constraint~\ref{eqh} prevents the formation of subtours by ensuring that all locations are connected in a single and continuous path.

\begin{figure*}[!t]
    \centering
    \includegraphics[scale=0.25, trim=0 100 0 100,clip]{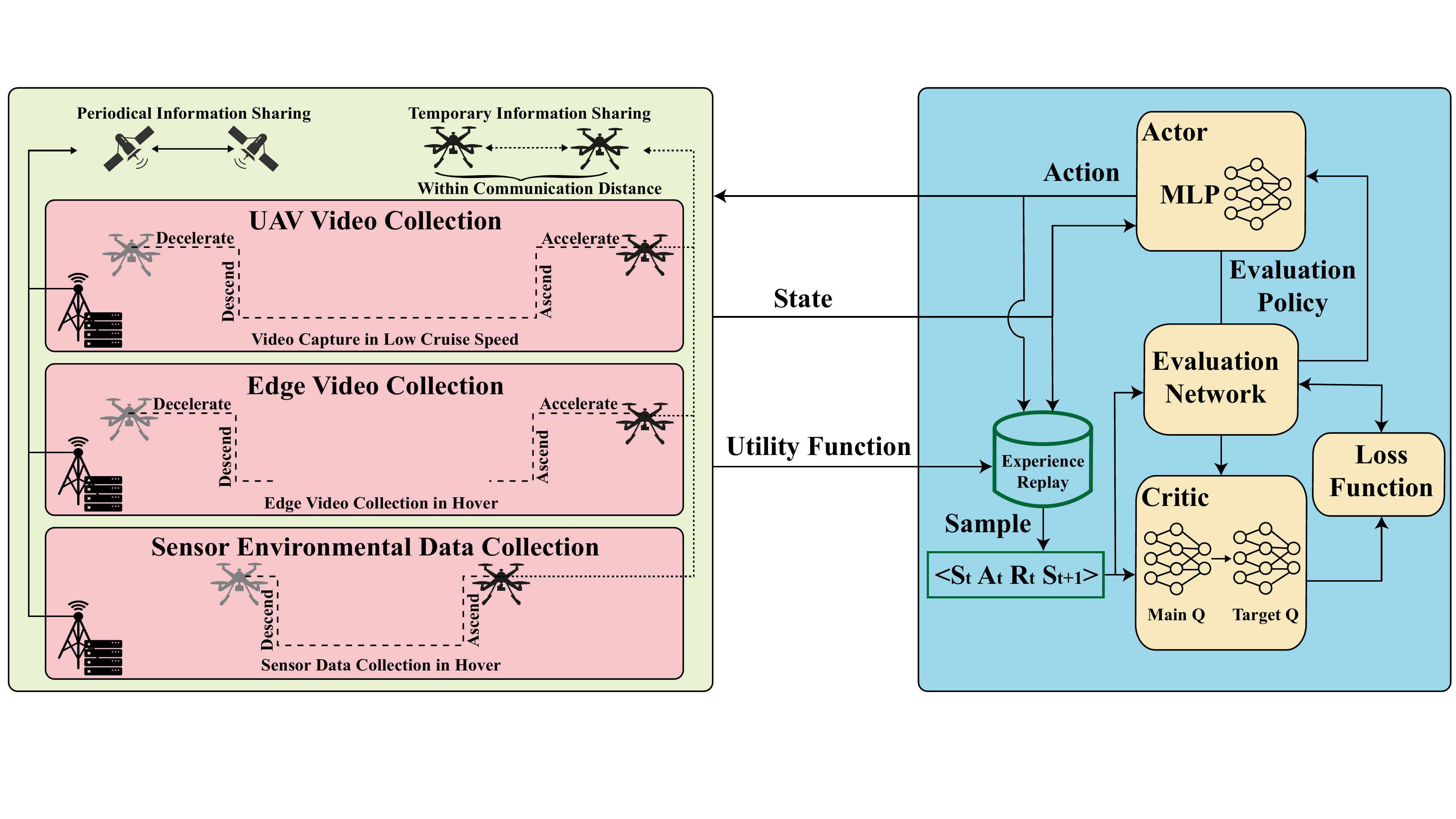} 
    \caption{Overview of the proposed framework \sysname.}
    \label{fig:Overview of the MASAC.}
\end{figure*}

\subsection{Curved Distance Metric}~\label{sec:cdm}

To incorporate realistic flight paths, we introduce a curved distance metric \( \tilde{d}_{nm} \) between each pair of tasks \( n \) and \( m \). This curved distance is expressed as:
\begin{equation}
\tilde{d}_{n,m} = 
\begin{cases} 
      d_{n,m}, & \text{if } \theta_{n,m} = 0, \\[10pt]
      \frac{d_{n,m} \cdot \theta_{n,m}}{2 \sin\left(\frac{\theta_{n,m}}{2}\right)}, & \textit{otherwise}
\end{cases}
\end{equation}
where \( d_{n,m} \) is the Euclidean distance between the locations of tasks \( n \) and \( m \). The central angle is a function of the task types at locations \( n \) and \( m \), defined as:
\begin{equation}
     \theta_{n,m} = \max\left(0, \ln \left(\frac{C_n S_n}{C_m S_m}\right)-1\right) 
\end{equation}
where \( C_n \) and \( C_m \) are the computational requirements for tasks \( n \) and \( m \), \( S_n \) and \( S_m \) are the storage requirements for tasks \( n \) and \( m \). This formulation reflects the need for UAVs to follow curved paths based on mission-specific parameters, which ensure the curved distance \( \tilde{d}_{n,m} \) is always longer or equal to the direct Euclidean distance.

This problem is a non-Euclidean TSP problem, which integrates storage constraints directly into the distance metric to avoid the limitations of the traditional Euclidean TSP. In a Euclidean setup, routes are optimized by straight-line distances, often overlooking storage constraints and requiring additional checks to ensure feasibility. By embedding storage into a non-Euclidean formulation, we avoid these post-solution checks, as storage limitations are factored into the routing decisions upfront. Since solving a non-Euclidean TSP is computationally complex, we address this issue with an algorithm that utilizes enhanced distributed actor-critic networks, which efficiently achieves both route feasibility and optimized performance.

\subsection{Design of \sysname}

We propose a multi-task video processing framework based on enhanced distributed
actor-critic networks to obtain satisfactory service performance. \figurename~\ref{fig:Overview of the MASAC.} shows the overview of our proposed framework. In \sysname, each ground station acts as an agent that allocates tasks to the UAVs under its control based on the current information available, including the UAVs' battery levels, task locations, and task allocation status. The objective is to maximize task collection and processing rates while minimizing energy consumption. As illustrated in \figurename~\ref{fig:Overview of the MASAC.}, upon departing from their respective ground stations, each UAV collects and processes three types of tasks: UAV video collection, edge video collection, and sensor environmental data collection. These tasks with different priority levels are allocated the corresponding task processing actions specified in Section~\ref{sec:tpm}.


\shen{A distributed algorithm is essential for \sysname, since centralized methods relying on information synchronization encounter challenges due to high latency and costly satellite communication in resource-constrained infrastructure-less environments. This makes it impractical for ground stations to synchronize every processing step. Through distributed decision-making, each ground station autonomously assigns tasks to UAVs, adapting to local conditions dynamically to improve service reliability. However, independent task allocation may lead to duplicate assignments among UAVs managed by different ground stations, resulting in inefficient use of UAV resources. To mitigate this issue, UAVs exchange information about their ground stations when they enter the communication range during flight (\textbf{temporary information sharing}), reducing duplicate task assignments and improving resource utilization. Additionally, all ground stations periodically upload information to the satellite after every time step $T_0$ (\textbf{periodical information sharing}), enabling access to task and resource information from other ground stations, thereby further enhancing task collection and processing efficiency. Upon completing task collection, UAVs return to their initial ground stations for recharging, ensuring continuity in mission operations.}

We transform the aforementioned SMD-MTSP into a Markov Decision Process (MDP), four ground stations at the corners of the \( A \times A \) region, with each station functioning as an individual agent. These agents make optimal decisions based on local observations to maximize cumulative rewards. Other important elements in our designed algorithm are outlined as follows:

\subsubsection{State space}
The environmental state $s_t$ includes the battery status of all UAVs and the locations of task sites at time  $t$, as well as the task allocation status.

\subsubsection{Observation space}
At time $t$, each agent $i$  can only access its local observation $o^t_i$, which does not provide a complete view of the system state. As shown in ~\figurename~\ref{fig:Overview of the MASAC.}, in the absence of information sharing, each agent can only access the current battery status of the UAVs at the ground station, the locations of task sites, and partial task allocation status.

\subsubsection{Action space}
At time $t$, each agent $i$  selects an action $ a^t_i$ for the next period, which involves the ground station allocating tasks to the UAVs and optimizing their action paths based on the current battery status of the UAVs and the information regarding task locations.

\subsubsection{Transition probability}
$\mathcal{P}$: $p(s_{t+1}|s_t,a_t)$ is the probability of transition from state $s_t$ to $s_{t+1}$ when all the agents take joint action $a_t$ = \{$a_{i,t}\}_{i=1}^N \in \mathcal{A}$.

\subsubsection{Reward function}
Agent $i$ obtains a reward $r_i^t$ by reward function $\mathcal{S}\times\mathcal{A}_1\times \cdots \times\mathcal{A}_N\rightarrow \mathcal{R}$. In this paper, the objective of the agents is to maximize the task collection rate and the task processing rate, while reducing the energy consumption associated with task processing. The reward function is formulated based on the optimization objective defined as follows:
\begin{equation} \label{eq:reward}
r^t = \sum_{\forall m} C_t \cdot \left( \frac{\mu \cdot \left( e^{\omega \cdot \textit{CR}_m} - 1 \right)}{3} \cdot Pr_m^{CT} - \epsilon \cdot E_{\text{process},m} \right) \cdot \Phi
\end{equation}
where $\textit{CR}_m$ is the completion rate of task $m$, $Pr_m^{CT}$ is the priority of task $m$ (See Section~\ref{subsec:nwmodel}), $E_{\text{process},m}$ is the energy consumed by processing task $m$, $C_t$ is the task collection rate at time $t$, while  $\mu$, $\omega$, $\epsilon$ and $\Phi$ are hyperparameters.

\subsubsection{Policy} 
At time  $t$, agent $i$  chooses an action $a_t $according to a specific policy $\pi_t$  to maximize the total reward. 

Each agent determines the optimal policy during the policy update process, with the objective of maximizing cumulative rewards while also making the policy more stochastic by:
\begin{equation}
    \pi^{*} = \arg\max_{\pi}\mathbb{E}_{\pi}\left(\sum_{\forall t}r(s_t,a_t)+\alpha_H H(\pi_t(\cdot|s_t))\right)
\end{equation}
where $\alpha_H$ is a trade-off coefficient, and $H(\pi_t(\cdot|s_t))$ is the entropy calculated by:
\begin{equation}
    H(\pi_t(\cdot|s_t))=-\sum_{a}\pi_t(a_t|s_t)\log \left(\pi_t(a_t|s_t)\right)
\end{equation}

During each training step, a subset of data is randomly sampled from the replay buffer in order to update the parameters of both the actor networks and critic networks. 
The target for Q functions is expressed by:
\begin{equation}
      \begin{split}
       y_i &=  r_i + \\&\gamma \mathbb{E}\left(\min_{i=1,2}{Q_{\phi_{\text{targ},i}}{(s_{t+1},\tilde{a}_{t+1})}} - \alpha \log \pi_{\theta}(\tilde{a}_{t+1}| s_{t+1})\right),\\& \tilde{a}_{t+1} \sim \pi_{\theta}(\cdot| s_{t+1})
       \label{eq:target}
       \end{split}
       \end{equation}

Based on the obtained targets, the critic networks are updated by minimizing the loss function defined as:
\begin{equation}
    \mathcal{L} =
    \mathbb{E}_{s_t,a_t,r_t,s_{t+1}}\left(Q_{\theta,i}(s_t,a_1,...,a_n)-y_i\right)^2
    \label{eq:loss}
\end{equation}

Then, the current actor network is updated using the following loss function:
\begin{equation}
    \mathcal{L_\pi(\theta)} = \frac{1}{N} \sum_{i=1}^{N} \left(
\alpha \log \pi_{\theta}(\tilde{a}_{t}| s_{t})  -
\min_{i=1,2}Q_{\theta,i}(s_t,\tilde{a}_t) \right)
\label{eq:actor loss}
\end{equation}



\begin{algorithm}[tb!]
\caption{Collaborative UAV Task Allocation and Information Sharing Based on Enhanced Actor-Critic Networks}
\label{algo:uav}
\begin{algorithmic}[1]

\STATE \textbf{Input:} $\mathcal{I}(T-1)$: UAV information at time $T-1$ for all ground stations.
\STATE \textbf{Output:} $\mathcal{I}(T)$: UAV information at time $T$ for all ground stations.

\FOR{each ground station $g$}
    \STATE Initialize $\pi_{\theta_g}$ and $Q_{\theta_g}$.
    \STATE Observe state $s_g(T-1)$, select action $a_g(T)$, execute and record $r_g(T)$.
    \STATE Update $Q_{\theta_g}$ by Equation~\ref{eq:loss}
    and $\pi_{\theta_g}$ by Equation~\ref{eq:actor loss}.
    
    \FOR{each nearby UAV $j$ such that $d(i,j) \leq d_\text{threshold}$ and $i, j$ from different stations}
        \STATE UAVs $i$ and $j$ share information, and update $\mathcal{I}(T)$.
    \ENDFOR
    \IF{$T$ MOD $T_0$ is 0}
            \STATE Conduct information sharing among all stations. \STATE Update $\mathcal{I}(T)$.
        \ENDIF

    \FOR{each ground station $g$}
        \IF{information from station $n$ is missing at station $g$}
            \STATE Estimate $\mathcal{I}_n(T)$ using decay model with most recent UAV information update time $T_\textit{recent}$: 
            \[
            \mathcal{I}_n(T) = \mathcal{I}_n(T-1) \cdot e^{-\lambda (T - T_\textit{recent})}.
            \]
        \ENDIF
        \STATE Update $\mathcal{I}(T)$ for station $g$.
    \ENDFOR
\ENDFOR

\end{algorithmic}
\end{algorithm}

Algorithm~\ref{algo:uav} shows the details of the distributed algorithm we developed, which consists of three main phases: experience collection, network training, and information sharing. Lines 3 to 5 represent the experience collection and network training processes. Lines 6 to 8 indicate that when UAVs from different ground stations are flying close to each other while executing tasks, they can communicate and share information including the status of UAVs battery usage and the task allocated between the two ground stations. To prevent situations where UAVs from different ground stations are unable to meet, we establish a schedule to share information from all ground stations via \shen{satellite communication} every $T_0$ time steps, as shown in lines 9 to 11 in  Algorithm~\ref{algo:uav}. In the absence of shared information, each ground station predicts the information of other ground stations based on a time-decaying model.

To calculate the expected performance gap of the distributed actor-critic networks algorithm with \( K \) ground stations and a per-step communication probability \( p \), we consider the likelihood of each station missing updates from all \( K-1 \) other stations. For any given step, the probability that a station does not receive information from a specific other station is \( 1 - p \), and the probability of missing information from all \( K-1 \) stations in a single step is \( (1 - p)^{K-1} \). Over a $T_0$-step interval, the probability of not receiving any updates from any other station is \( (1 - p)^{T_0 \cdot (K-1)} \), leading to the worst-case accumulated decay error:
\begin{equation}
\Delta V_{\text{worst}} = \mathcal{I}_n(0) \cdot e^{-\lambda \cdot T_0}
\end{equation}

To capture the expected performance gap across possible update scenarios, we weight the error based on the probability of receiving at most one update within  $T_0$ steps. The expected gap is given by:
\begin{equation}
\begin{split}
\mathbb{E}(\Delta V) = & \sum_{n_\text{step}=0}^{T_0} \binom{T_0}{n_\text{step}} \cdot [(1-p)^{K-1}]^{n_\text{step}} \cdot \\& [1 - (1-p)^{K-1}]^{T_0 - n_\text{step}} \cdot \mathcal{I}_n(0) \cdot e^{-\lambda \cdot n_\text{step}}
\end{split}
\end{equation}

In the best-case scenario, where updates are received at each step, the gap remains zero (\( \Delta V_{\text{best}} = 0 \)), while the expected gap reflects the weighted accumulation of decay errors based on the probability of missing information over the interval.

%% file: 6_experiment.tex
\section{Experiment}\label{sec:evaluation}

\subsection{Experimental Setup}
We build a platform that simulates the aerial computing network for performance evaluation on \sysname. In this platform, UAVs take off from airports located at the four corners to collect and process three different types of tasks mentioned in Section~\ref{sec:tpm} within an $A \times A$ region. The three types of mission data include UAV video collection, edge video collection, and sensor environmental data collection. UAV video data typically consists of high-resolution recordings, each lasting between 1 to 2 minutes, with file sizes ranging from 100 to 200 MB per segment. Edge video data primarily includes a mix of short video clips and images, with a cumulative size of approximately 300 to 500 MB per data transfer~\cite{zheng2020realizing}. In contrast, the sensor data includes environmental metrics and other situational parameters, and is relatively lightweight, with file sizes between 1 and 10 MB per dataset. When the UAVs return to the airport, they transmit the collected data to the ground station and ultimately return to the original airport to recharge, ensuring sufficient battery for the next time period’s tasks. Within the $A \times A$ region, 110 task locations are randomly distributed, and in each experimental round, the task locations change, with varying densities to better simulate real-world scenarios. 
To simulate the video processing capability of the UAVs, we test the processing time for videos of different lengths using the YOLOv5 algorithm on a Raspberry Pi 4 Model B. Other main parameters used in the experiment are
summarized in Table~\ref{tab:params}.

\begin{table}[tb]
 \caption{Main experimental parameters}
 \label{tab:params}
 \centering
\begin{tabular}{ll}
   \hline
   Parameter  & Value \\
   \hline \hline
   Number of UAVs& 16\\
   Number of tasks $|Tk|$ & 110, 150, 170 \\
   Area size ($A \times A$) & 1500~m $\times$ 1500~m \\
   Minimum height of UAVs $z_{min}$ 
   & 50~m~\cite{HaoTMC24}\\
   Maximum height of UAVs $z_{max}$& 100~m~\cite{HaoTMC24}\\
   Channel bandwidth of UAV to Edge $B^G$ & 10~MHz~\cite{DaiTMC24} \\
    Channel bandwidth of UAV to UAV $B^U$ & 40~MHz~\cite{DaiTMC24} \\
   Noise power $N^{U2G}_0$, $N^{U2U}_0$  & -100~dBm~\cite{HaoTMC24}\\
    Computation capacity of UAVs  & [10,20] GigaCycles~\cite{DaiTMC24} \\
   Transmit power of UAV $P_i$ & 5~W~\cite{HaoTMC24} \\
   Actor-network learning rate & $0.001$ \\
   Critic-network learning rate & $0.001$ \\
   Soft update parameters  &  $0.005$ \\
   Discount factor  & $0.99$ \\
   Batch size & 64 \\
   Number of agent & 4 \\
   Interval of information sharing $T_0$ & 10\\
   \hline
  \end{tabular}
\end{table}

We verify the effectiveness of our proposed \textbf{\sysname} against the following five comparative methods:\begin{itemize}
     \item \textbf{{Random (RND)}}: It randomly allocates the tasks to the available UAVs  while ensuring that each UAV can return to its designated ground station.

     \item \textbf{Genetic Algorithm {(GA)}~\cite{WenJCN22}}: A heuristic algorithm generates multiple task allocation strategies for each round of UAVs operations, ensuring that the UAVs can return to their destination. Each distinct task allocation strategy acts as an individual, and all individuals form a population. From this population, individuals with higher adaptability are bred to obtain the optimal task allocation scheme.

     \item \textbf{{Deep Q-Network (DQN)}}~\cite{BurhanuddinTVT22}: As a centralized method, deep neural networks (DNNs) are used to approximate the Q function, which evaluates the value of taking a specific action in a given state. DQN is a model-free reinforcement learning algorithm that learns the optimal policy by updating the Q values, thereby enhancing the overall task collection rate and task completion rate of the system.
     
     \item \textbf{{Multi-Agent Soft Actor-Critic (MASAC)}}~\cite{QinTWC23}: As a distributed method, each agent employs a policy gradient method and has its own actor and critic networks. Each agent independently updates its network and makes action decisions to maximize the total reward of the entire system.

     \item \textbf{Soft Actor-Critic {(SAC)}}~\cite{QinTWC23}: This is a centralized algorithm based on the policy gradient method, which learns the policy by maximizing the expected return. It utilizes two neural networks: the actor network and the critic network. The actor network is responsible for generating actions, specifically the task allocation strategies, by selecting the optimal task distribution based on the current states of all UAVs and the environment, along with the policy parameters. The critic network is employed to estimate the state value function and the state-action value function. It maximizes the utilization of the UAVs' battery while ensuring that the UAVs can return to their destination, thereby enabling the UAVs to collect and process as many tasks as possible.

\end{itemize}

To ensure a fair comparison, all DRL-based baselines utilize the same reward function as our proposed \sysname, while the values of hyperparameter are consistent with those used in the original paper. It is worth noting that among the five comparison methods mentioned above, all except MASAC are centralized methods, which possess global information when allocating tasks to UAVs. As a result, the same task will not be redundantly assigned to different UAVs. In contrast, our proposed \sysname and MASAC, may lead to the same task being assigned to multiple UAVs, as each ground station does not have access to global information. Avoiding redundant task assignments is one of the problems that our proposal aims to address, and the effectiveness of our solution is summarized in the next subsection.

\subsection{Experimental Results}




\begin{figure*}[tb!]
	\centering
        \begin{minipage}[b]{.66\columnwidth}
		\centering
		\includegraphics[width=\columnwidth]{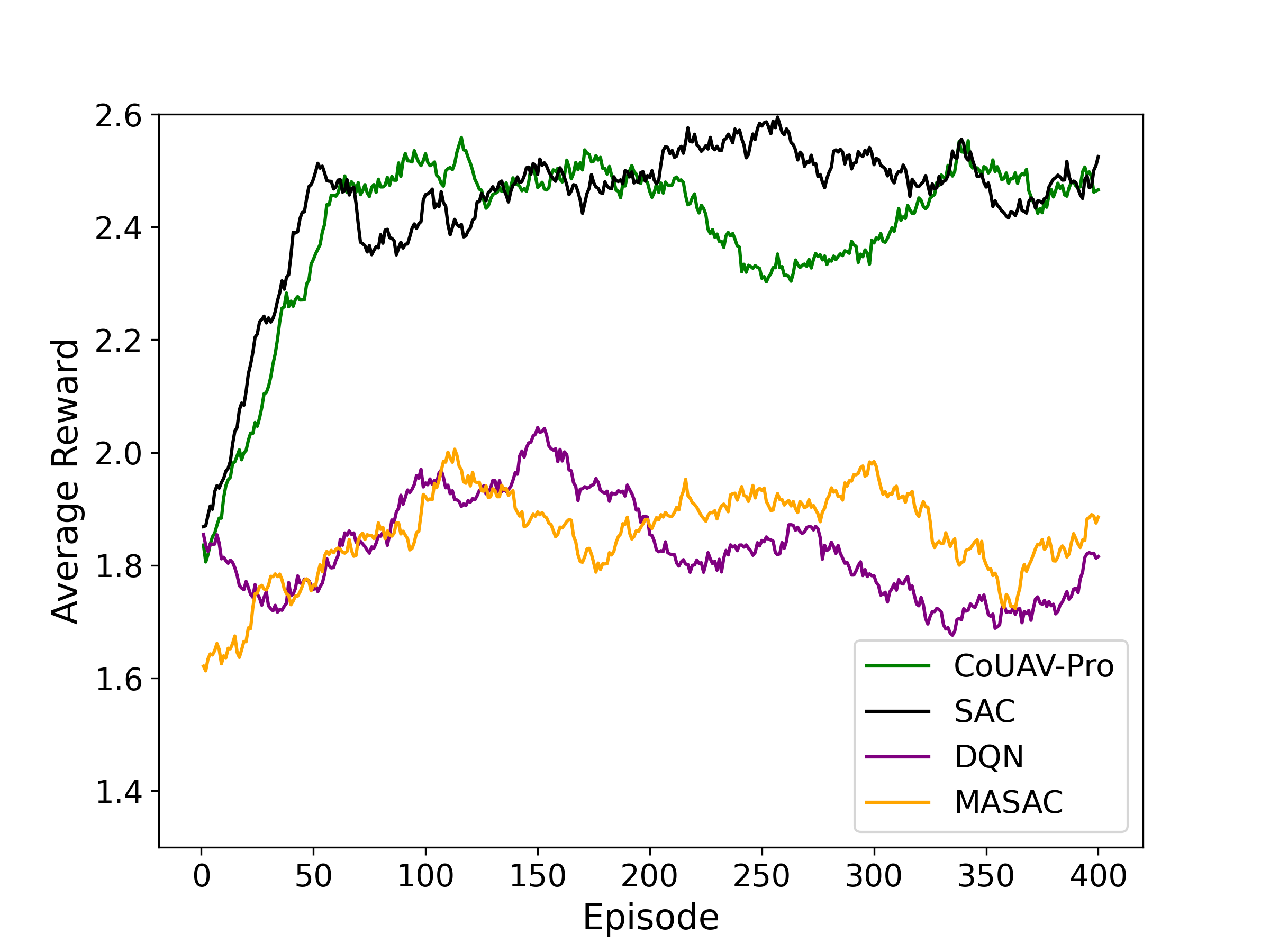}
		\subcaption{$|Tk|$ = 110 }\label{fig:reward_task110_train}
	\end{minipage}
        \begin{minipage}[b]{.66\columnwidth}
		\centering
		\includegraphics[width=\columnwidth]{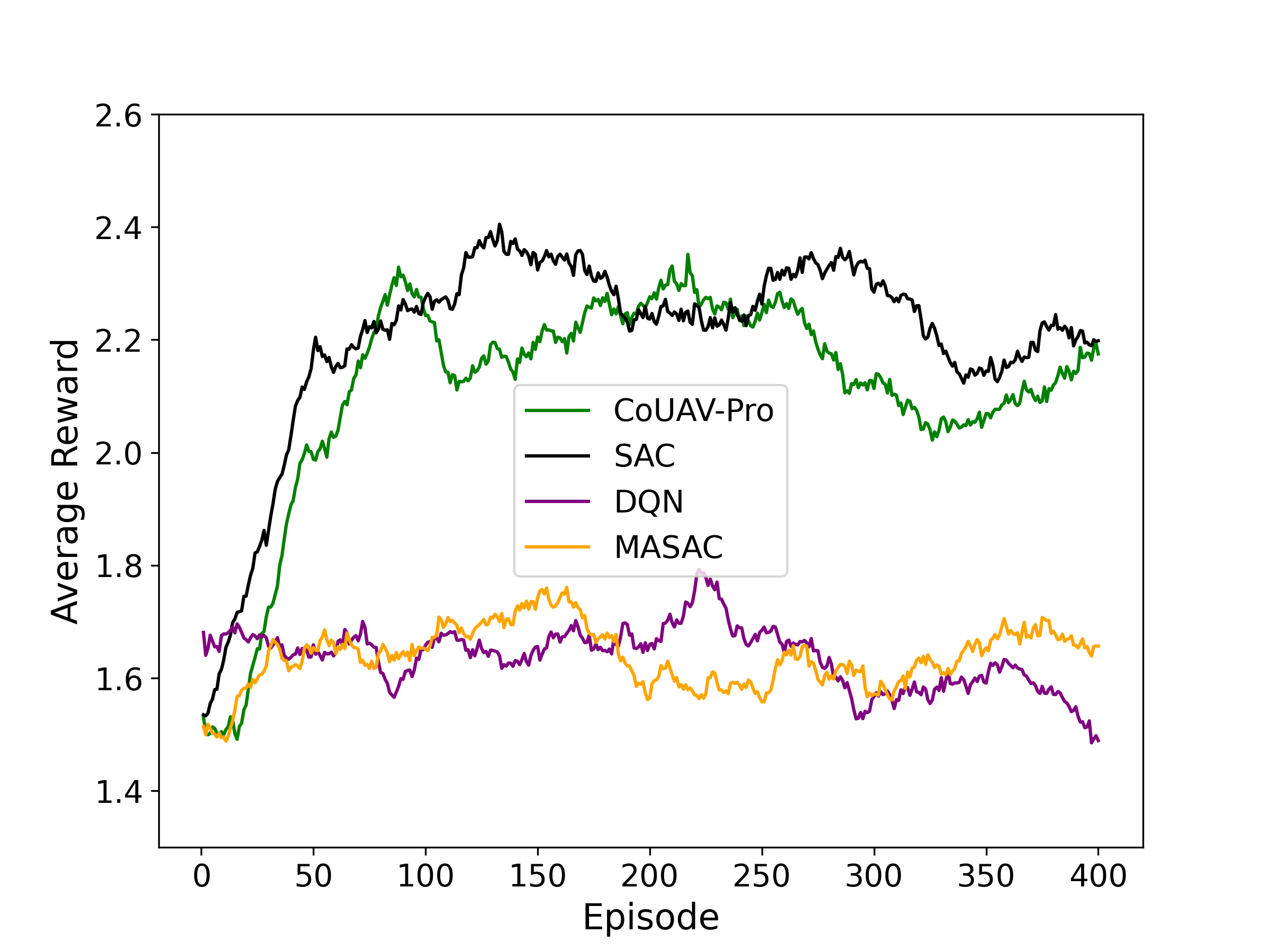}
		\subcaption{$|Tk|$ = 150 }\label{fig:reward_task150_train}
	\end{minipage}
	\begin{minipage}[b]{.66\columnwidth}
		\centering
		\includegraphics[width=\columnwidth]{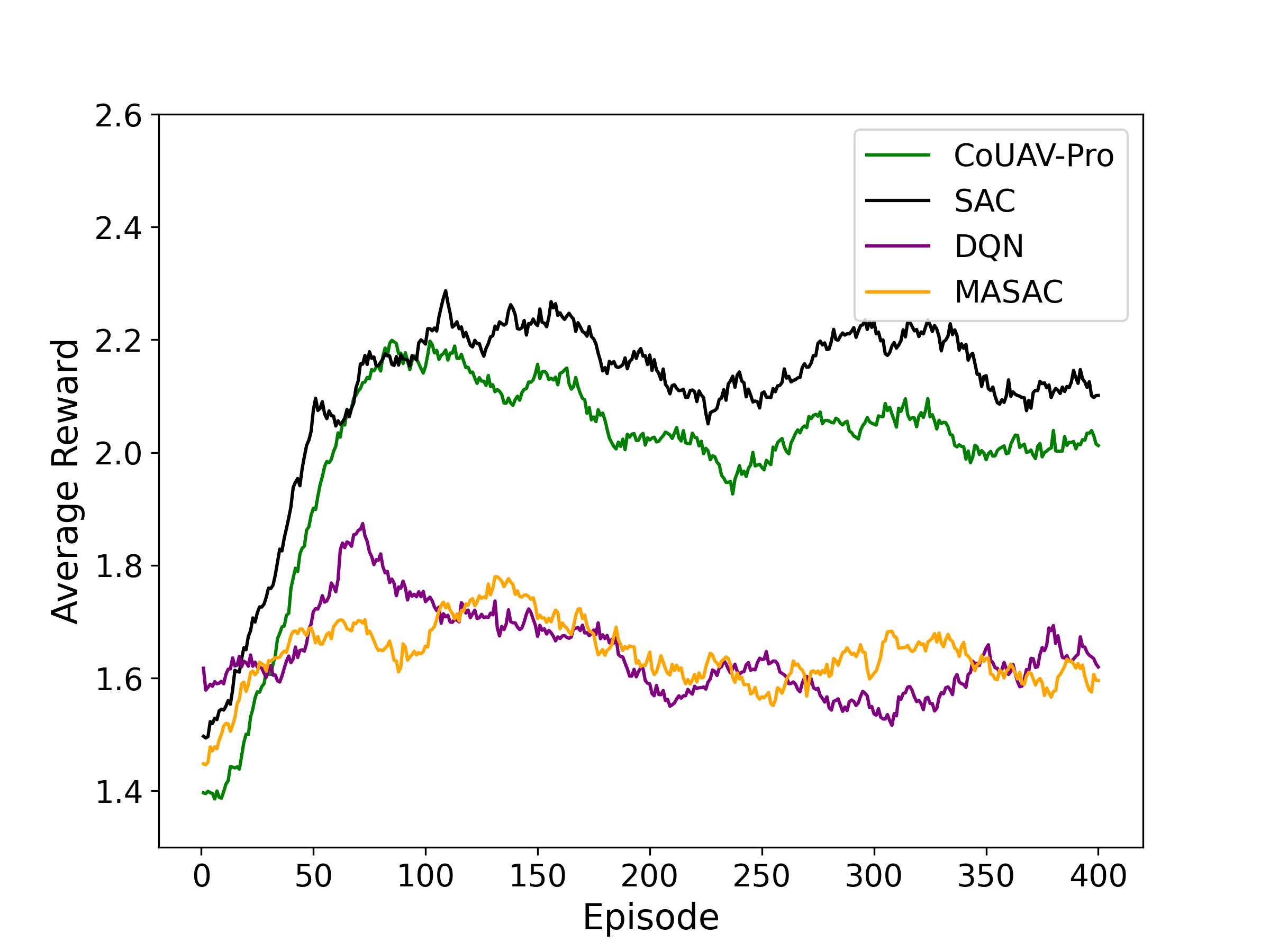}
		\subcaption{$|Tk|$ = 170 }\label{fig:reward_task170_train}
	\end{minipage}

	\caption{Convergence performance for DRL-based methods with various task volumes $|Tk|$. }
    \label{fig:convergence}
    \end{figure*}

\begin{figure*}[tb!]
	\centering
	\begin{minipage}[b]{.66\columnwidth}
		\centering
		\includegraphics[width=\columnwidth]{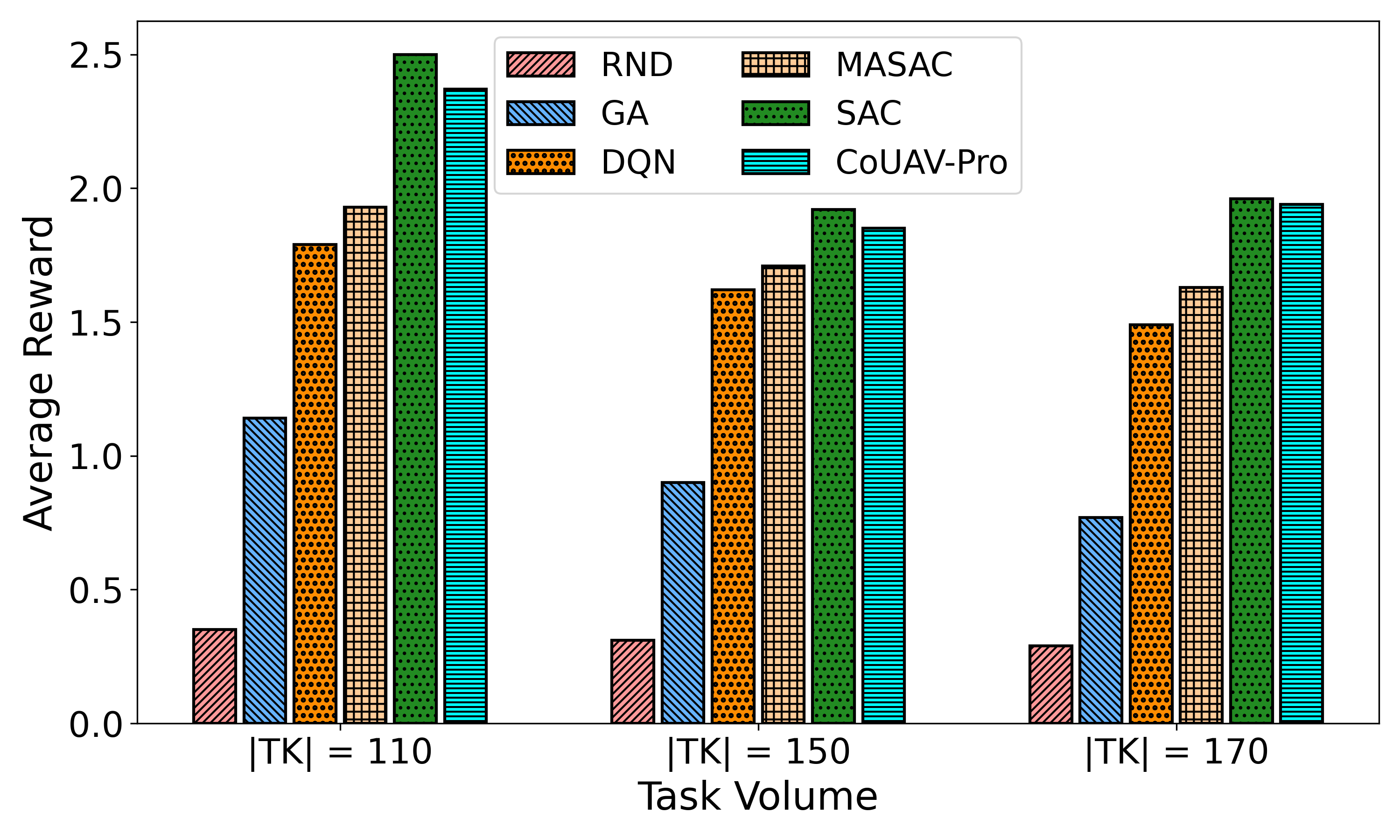}
		\subcaption{Average reward}\label{fig:reward}
	\end{minipage}
         \begin{minipage}[b]{.66\columnwidth}
		\centering
		\includegraphics[width=\columnwidth]{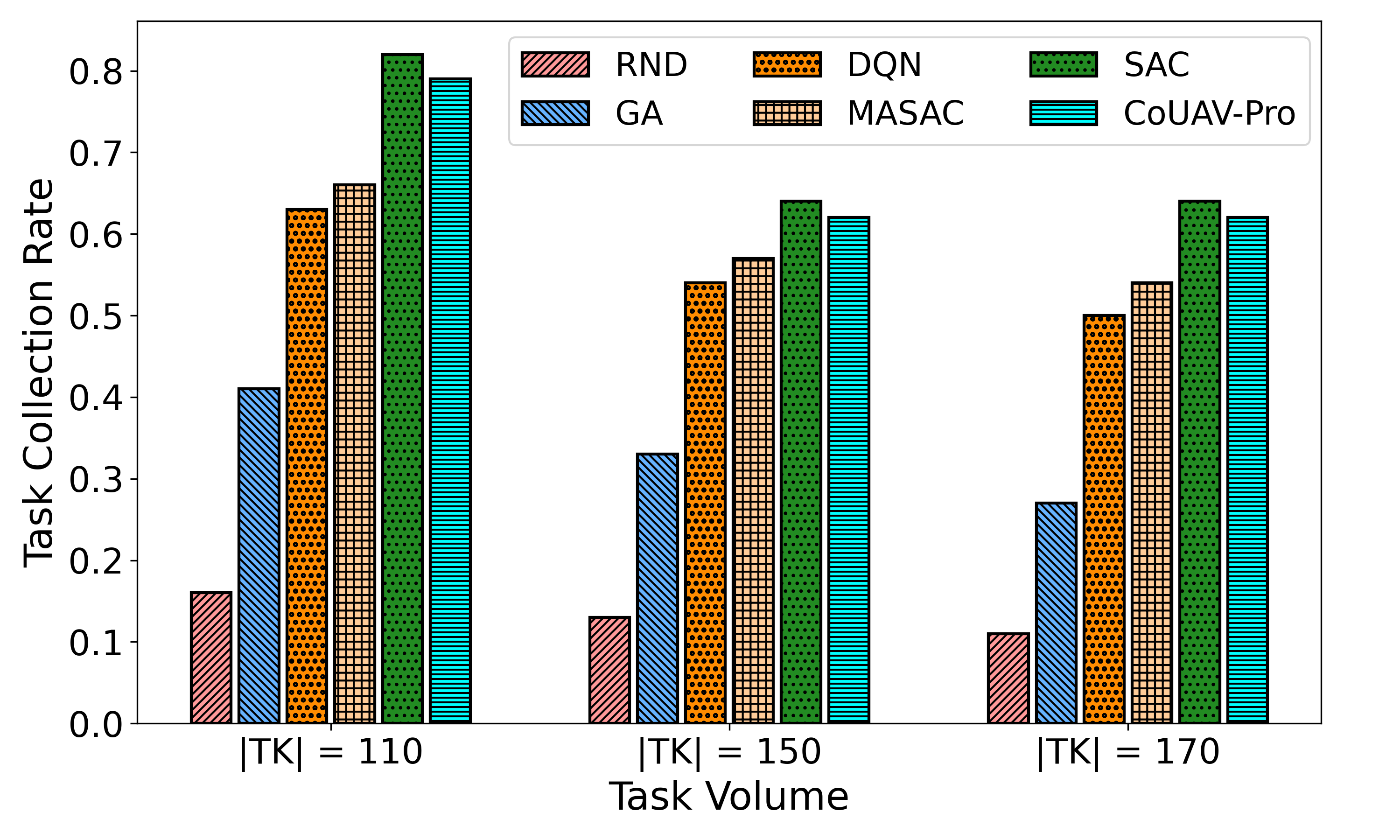}
		\subcaption{Task collection rate}\label{fig:select proportion}
	\end{minipage}
     \begin{minipage}[b]{.66\columnwidth}
		\centering
		\includegraphics[width=\columnwidth]{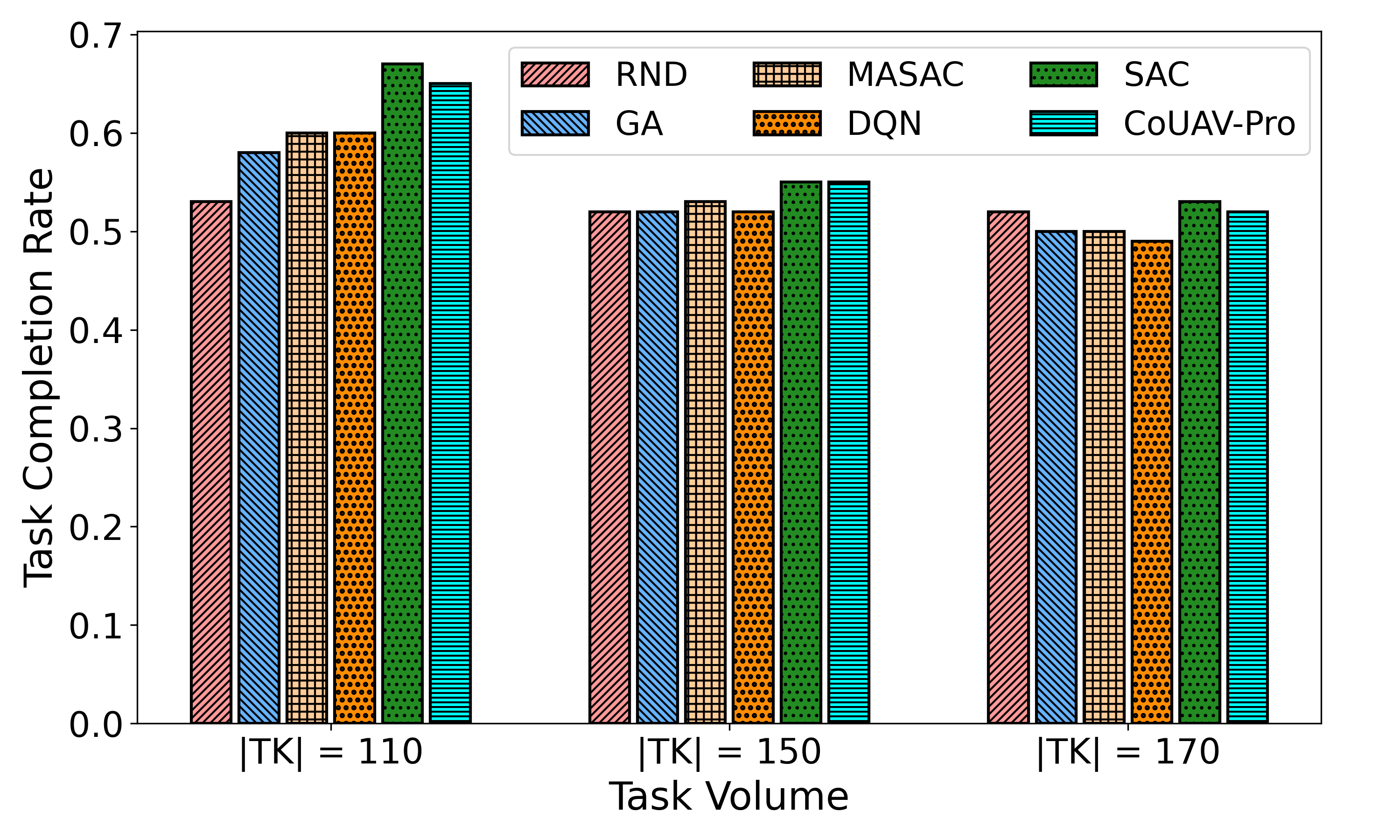}
		\subcaption{Task completion rate}\label{fig:completion}
	\end{minipage}\newline 
        \begin{minipage}[b]{.66\columnwidth}
		\centering
		\includegraphics[width=\columnwidth]{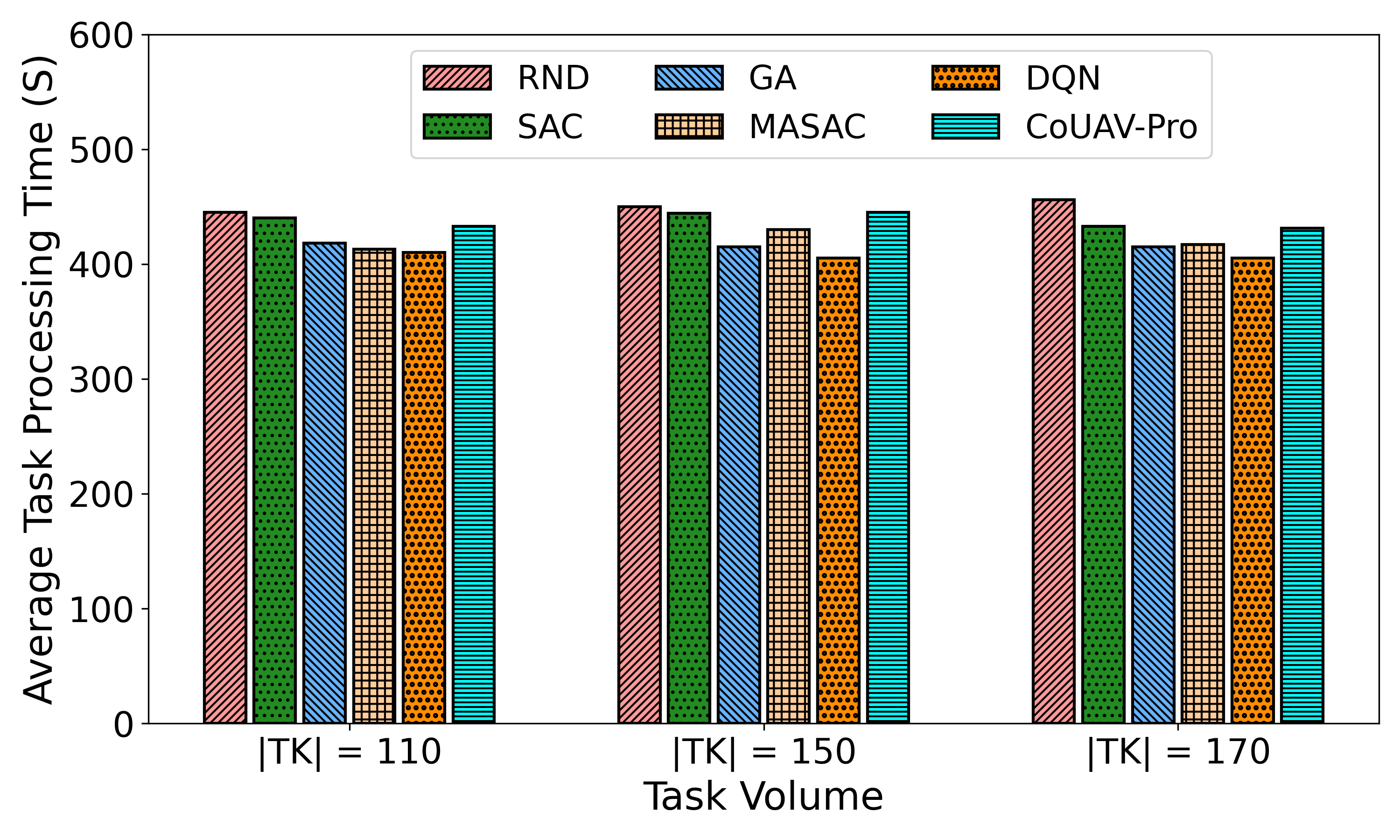}
		\subcaption{Average task processing time }\label{fig:delay}
	\end{minipage}
	\begin{minipage}[b]{.66\columnwidth}
		\centering
		\includegraphics[width=\columnwidth]{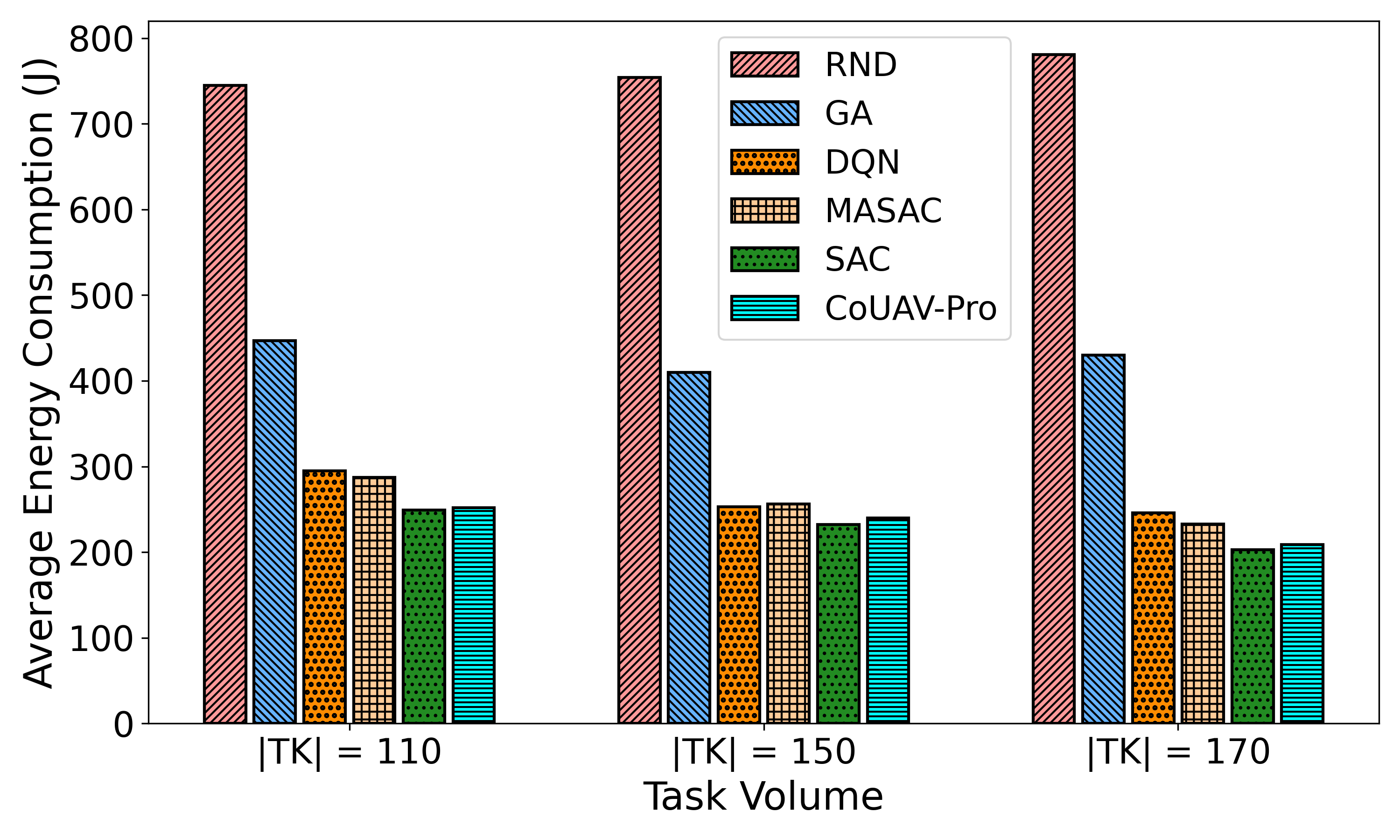}
		\subcaption{Average energy consumption}\label{fig:energy}
	\end{minipage}  
	\begin{minipage}[b]{.66\columnwidth}
		\centering
		\includegraphics[width=\columnwidth]{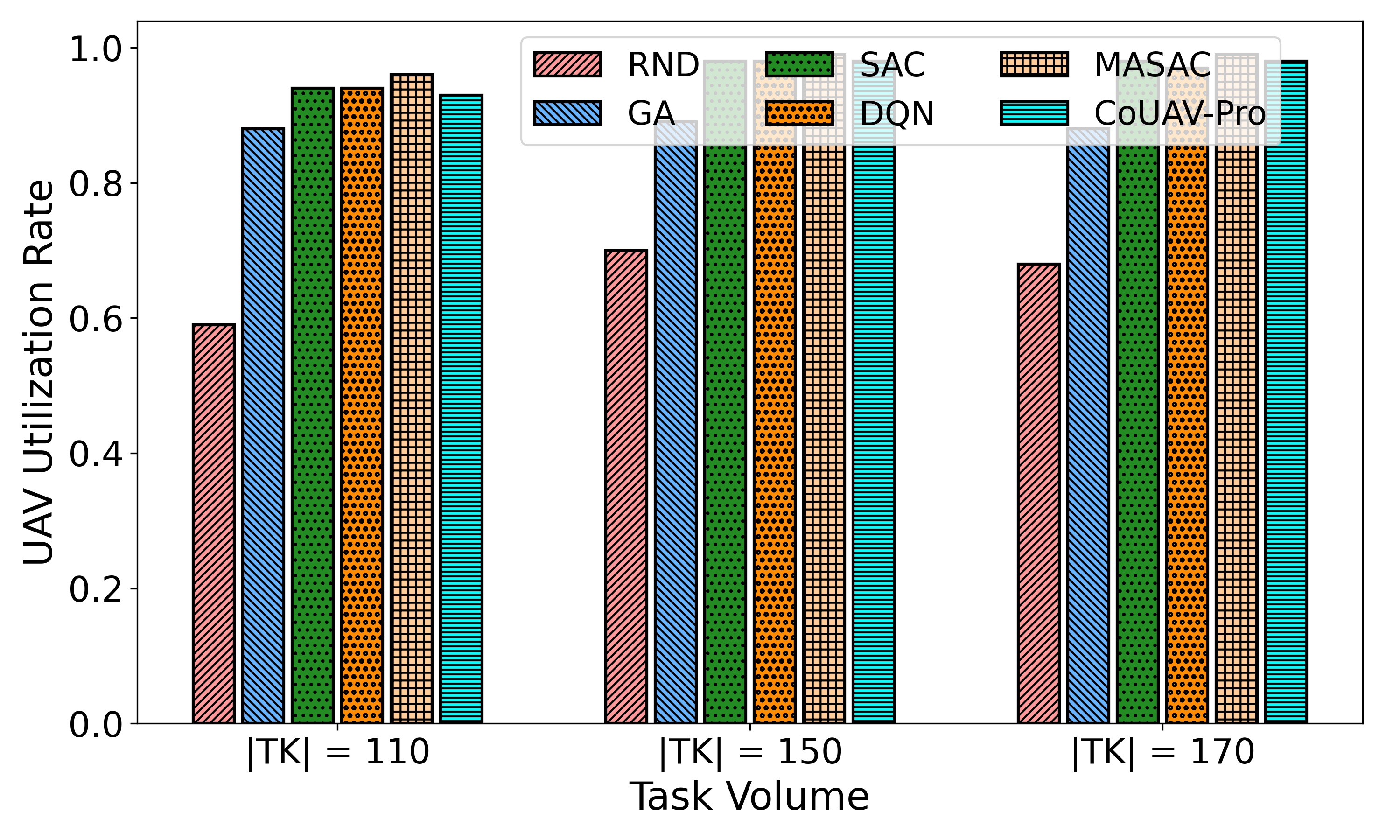}
		\subcaption{UAV utilization rate}\label{fig:uav used}
	\end{minipage}
	\caption{Overall task processing performance of different methods.}
 
        \label{fig:overall performance}
\end{figure*}

First, we evaluate the training convergence performance of various DRL-based methods under different task scales. We set the number of tasks to be processed in each round to 110, 150, and 170, respectively, with a batch size of 64 for all methods during the training process. 

\figurename~\ref{fig:convergence} illustrates the convergence performance in terms of reward obtained by four methods: our proposed \sysname, SAC, DQN, and MASAC.  Here, the reward is associated with the task collection rate, task completion rate, and energy consumption, as defined in Equation~\ref{eq:reward}. It can be observed that as the scale of tasks increases, the overall reward decreases. This is because with a fixed number of UAVs available for processing tasks, an increase in the scale of tasks results in a decrease in both the rates of task collection and completion.  The results indicate that as training progresses, our proposed \sysname achieves a performance that approximates those of the centralized SAC method. Due to its relatively inflexible policy adjustments, DQN fails to explore effectively in complex environments and quickly reaches stability. Conversely, as a distributed method, MASAC suffers from unsatisfactory reward prefetching since communication between ground stations is not possible. After stabilizing at 400 training episodes, the average task reward for \sysname, under the maximum task scale 170, is 2.05, while the average task rewards for the other methods are as follows: SAC (2.11), MASAC (1.69), and DQN (1.60).

These results underscore the importance of the shared mechanism inherent in our proposed method \sysname, which leverages communication between UAVs and periodic information sharing among all ground stations to facilitate collaborative training aimed at maximizing reward value. For all methods, we saved the model obtained after 400 episodes for the remaining experiments.

Next, we evaluate the policy performance of different methods based on various criteria. \figurename~\ref{fig:overall performance} shows the performance of each method under six criteria for three different task scales. Compared to four centralized methods (SAC, DQN, GA and RND) and one distributed method (MASAC), our proposed \sysname generally approaches the performance of the centralized SAC, while outperforming DQN, MASAC, GA and RND in terms of all six criteria.

In general, although the centralized RND method possesses global information, its task allocation for UAVs is conducted randomly, which may lead to over-utilization of UAV resources. This limitation constrains the resource selection for future computational tasks. The centralized GA method utilizes global information and simulates the process of biological evolution through mechanisms like selection, crossover, and mutation. This method gradually enhances problem solutions, leading to more efficient task allocation strategies than those achieved through the random selection method RND. However, as a heuristic search method, GA does not guarantee the discovery of a global optimum, particularly when faced with highly complex and nonlinear problems. Meanwhile, although the centralized DQN method has access to global information, it indirectly influences the policy through updates to action values. While this method can be effective in simpler environments, it lacks the flexibility needed for policy adjustment in more complex scenarios. MASAC is also a distributed method like our proposal; however, due to the relative independence and lack of communication among ground stations, there is a tendency for resource wastage by UAVs when the task scale is relatively small, leading to performance deterioration.

Our proposed \sysname aims to determine optimal task allocation by analyzing factors such as task completion rates and task acquisition rates. Unlike the centralized SAC that operates with a single agent possessing global information, \sysname consists of four agents that only have access to their own information. These agents communicate through the UAVs they managed to facilitate more efficient task allocation, thereby avoiding the execution of the same task by different UAVs. Additionally, the agents share their respective information at specified time intervals to collect data from the network and tasks, enabling optimal decision-making training in the resource and task allocation process.

In details, \figurename~\ref{fig:reward} shows the average task rewards of our method exceed those of MASAC, DQN, GA, and RND across three different task scales, ranking only behind SAC. This improvement is attributed to the shared mechanism introduced in \sysname, which allows our proposal to significantly outperform the other methods in average task reward when the task scale is 110. In contrast to MASAC, our method not only incorporates the observations and actions of the agents themselves, but also integrates the observations and actions of all other agents at specified time intervals. This enables more informed action selection based on the current state, resulting in higher returns. 

In terms of task collection rates shown in \figurename~\ref{fig:select proportion}, as the task scale increases, the task collection rates of all methods decrease due to the number of UAVs is fixed. SAC exhibits the highest task collection rate, while our proposed method approaches the performance of SAC. Both MASAC and DQN achieve task collection rates around $60\%$, with MASAC outperforming DQN. As the scale of tasks increases, the likelihood of duplicate task assignments to UAVs from different ground stations decreases with the distributed method, leading to better performance of MASAC compared to centralized method DQN.

Regarding task completion rates as shown in \figurename~\ref{fig:completion}, our proposed method \sysname achieves performance that approximates that of SAC method despite variations in task scale, while outperforming other comparative methods. This effect is particularly evident when the task scale is 110. 
In contrast, for task scales of 150 and 170, the task completion rates across various tasks remain relatively consistent. This can be attributed to the UAVs' inability to process a high volume of tasks efficiently as the task scale increases.

From the perspective of average task processing time (\figurename~\ref{fig:delay}) that indicates the time from when all UAVs take off until the last UAV returns to the endpoint, our proposed method, along with SAC, results in a slightly longer return time for the last UAV compared to other reinforcement learning methods due to higher task collection and completion rates. In contrast, GA has a lower collection rate, allowing UAVs to return to the endpoint more quickly. In terms of energy consumption shows in \figurename~\ref{fig:energy}, there is minimal variation in energy consumption among the reinforcement learning methods. SAC exhibts demonstrates the lowest energy consumption, closely followed by our proposed \sysname. However, GA and RND have higher energy consumption due to the complexity of situation changes. Overall, our proposed \sysname not only achieves lower task processing times but also outperforms MASAC, DQN, GA, and RND in terms of task collection rate, task completion rate, energy consumption, and rewards. By significantly reducing the frequency of communication between multiple agents, \sysname achieve satisfactory performance that approximates that of the centralized SAC method.

%% file: 7_conclusion.tex
\section{Conclusion}\label{sec:conclusion}
This paper develops a multi-task video processing solution empowered by enhanced distributed actor-critic networks within aerial networks in infrastructure-less environments. Our proposed \sysname optimizes task latency and energy consumption through the joint design of UAV flight trajectories, task offloading, and computational resource allocation. Our proposed \sysname enables efficient task distribution among multiple UAVs and facilitates information sharing during the data collection process, thereby enhancing service reliability and performance without a continuous reliance on a central node. The results indicate that \sysname achieves performance that closely approximates that of the centralized algorithm SAC in terms of task rewards, task processing energy consumption, task collection rate, and task completion rate, while significantly outperforming other comparative methods. This implies that \sysname minimizes high-cost communication with the central node, thereby reducing the overall energy consumption of the system and achieving robust, scalable performance to meet the demands of diverse CV tasks in infrastructure-less environments.